 \documentclass[twocolumn,prb,amssymb]{revtex4}
\usepackage{graphicx}
\usepackage{dcolumn}
\usepackage{bm}
\begin{document}
\title{Wetting in mixtures of colloids and excluded-volume
polymers from density functional theory}
\author{Pawe{\l} Bryk}
\affiliation{Department for the Modeling of Physico-Chemical Processes,
Maria Curie-Sk{\l}odowska University, 20-031 Lublin, Poland}
\email{pawel@paco.umcs.lublin.pl}
\date{\today}
\begin{abstract}
We use a microscopic density functional theory based on Wertheim's first
order thermodynamic perturbation theory to study wetting behavior of
athermal mixtures of colloids and excluded-volume polymers.
In opposition to the wetting behavior of the Asakura-Oosawa-Vrij model
we find the polymer-rich phase to wet a hard wall. The wetting
transition is of the first order and is accompanied by the prewetting transition. We do not
find any hints for the layering transitions in the partial wetting regime.
Our results resemble the wetting behavior in athermal polymer solutions.
We point out that an accurate, monomer-resolved
theory for colloid-polymer mixtures should
incorporate the correct scaling behavior in the dilute polymer regime
and an accurate description of the reference system.
\end{abstract}
\maketitle
\section{Introduction}
\label{sec:introduction}
Mixtures of nonadsorbing polymers and colloidal particles
often exhibit rich phase behavior \cite{Poon02,Tuinier03}. For certain polymer-to-colloid size ratios $q=R_g/R_C$
(where $R_g$ is the radius of gyration of the polymer and $R_C$ is the radius of colloid)
entropy-driven effective interactions may lead to stable colloidal gas, liquid and solid
phases even if all bare interactions are purely repulsive \cite{Gast83}. A simple
theoretical model giving an insight into this phenomenon
is the Asakura-Oosawa-Vrij (AOV) model of colloid-polymer mixtures
in which the polymers (modeled as spheres) are ideal and can overlap freely,
but the polymer-colloid and colloid-colloid
interactions are of the hard sphere type \cite{Asakura54,Vrij76}. Effective
attractive interactions in such systems arise
due to a tendency to decrease the volume excluded to the polymer coils \cite{Lekkerkerker92}.

The AOV model has attracted much attention due to its simplicity \cite{Brader03}, however 
when comparing to experiments the agreement is only qualitative. This
is due to the fact that the real polymers are nonideal.
The incorporation of the polymer nonideality can be tackled
at the monomer-resolved \cite{Fuchs02,Fuchs00,Shah03,Bolhuis02} or coarse-grained 
\cite{Warren95,Aarts02,Schmidt03} levels of description. 
The latter technique is particularly 
useful in the so-called ``colloid limit'' ($q\leq 1$),
in which the full mixture is well described by invoking the effective,
pairwise depletion potentials \cite{Moncho05}.
However, when the polymer dimensions are larger than the size of the colloid
particles, the resulting two-body effective potentials may be insufficient
to correctly describe the underlying mixture and the incorporation of the higher-order,
many-body terms becomes necessary \cite{Bolhuis03,Moncho03}. 

In addition to many industrial applications colloid-polymer mixtures 
offer a convenient tool to study important fundamental concepts
that are often difficult to investigate in simple fluids, such as
real space observation of the thermal capillary waves \cite{Aarts04a},
the capillary length \cite{Aarts05},
the interfacial width \cite{Hoog99b} and the interfacial tension of
the fluid-fluid interface \cite{Aarts03,Chen00,Chen01}.
Wijting {\it et al}. \cite{Wijting03a,Wijting03b}
studied the behavior of colloid-polymer mixtures close to a wall.
They found a wetting transition to a state in which the colloid-rich phase
wets completely a nonadsorbing planar hard wall. These findings qualitatively agree with the
results of density functional theory \cite{Brader02} and computer simulations 
\cite{Dijkstra02} carried out for the AOV model. 
Aarts {\it et al}. \cite{Aarts04b} used the Cahn-Fisher-Nakanishi approach to study
the wetting transition in mixtures of colloids and excluded volume polymers and
pointed out some subtleties associated with the precise measurements of the
contact angle \cite{Aarts04b,Aarts04c}.

Recently Paricaud {\it et al}. \cite{Paricaud03} examined the bulk phase behavior of 
mixtures of hard-sphere colloids and excluded-volume polymers within the
framework of Wertheim's first-order thermodynamic perturbation theory (TPT1) \cite{Wertheim87}.
They found a demixing transition into the colloid-rich and polymer-rich phases
provided that the ratio of the colloid diameter to polymer segment diameter is large enough.
The TPT1 description employs the microscopic description of the polymer,
consequently it is treated on equal footing with the colloid particles. 
This theory provides thus a straightforward procedure of taking the attractive/repulsive interactions 
into account. However, the Wertheim TPT1 approach is not free of its own deficiencies.
We note here that incorporation of the solid-fluid equilibrium into the framework,
while in principle possible \cite{Vega01},
is difficult to execute for the colloid-polymer mixtures.
Moreover, Boublik {\it al}.\cite{Boublik89,Boublik90} pointed out
that within the TPT1 theory the second virial coefficient scales quadratically with the chain length $M$
instead of $\propto M^{3\nu}$,
although it is possible to improve this deficiency \cite{Vega00}.
Finally, the value of the Flory exponent, $\nu$, resulting from the TPT1 approach
is 0.5 rather than 0.588 \cite{Paricaud03}. 
Despite these shortcomings, the TPT1 approach to colloid-polymer mixtures
provides a welcome departure from the well-studied AOV model.

In this work we focus on the surface phase behavior of model
colloid-polymer mixtures. We employ nonlocal density functional theory that was
used previously to investigate liquid-liquid interfaces of athermal mixtures of colloids
and excluded-volume polymers \cite{Bryk05}.
The density functional framework, proposed originally by
Yu and Wu  \cite{Yu02}, uses the TPT1 ideas and yields in the limit of bulk systems 
an equation of state identical to that from Ref.~\cite{Paricaud03}.
Since the DFT framework has already been described in detail \cite{Bryk05,Yu02} we recall only the basic
points of this approach.

\begin{figure}[t]
\includegraphics[clip,width=8cm]{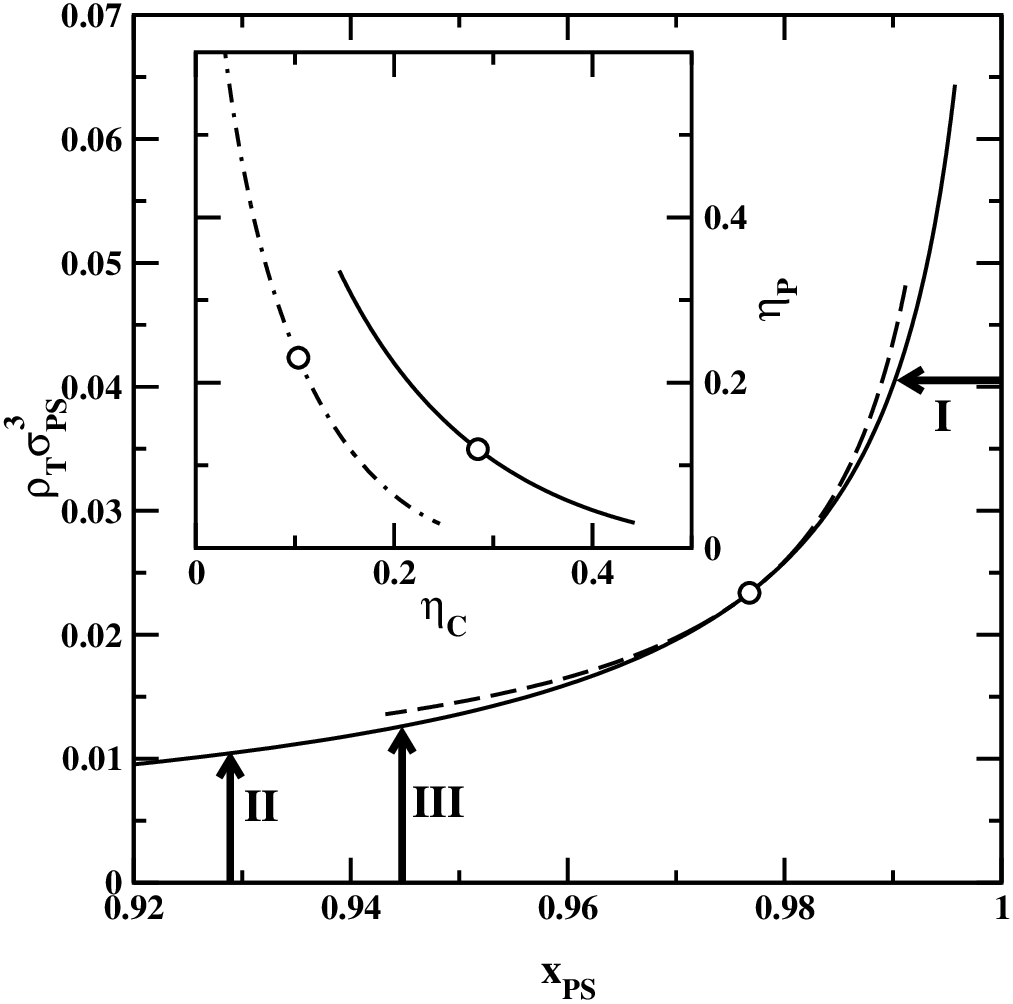}
\caption{\label{fig:1}
Bulk phase diagram for a colloid polymer mixture with $d=10$ and $M=100$.
The main figure shows the binodal (solid line), the spinodal (dashed line), and
the bulk critical point (open circle) evaluated
in the polymer segment mole fraction-total density plane.
The open circle denotes the bulk critical point.
The inset shows the binodals resulting from the TPT1 (solid line) and free volume (dash-dotted line)
theories evaluated in the colloid-polymer packing fraction representation.
}
\end{figure}

\section{Theory}
\label{sec:theory}
We model the colloids as hard spheres of diameter $\sigma_C$
and the polymers as chains composed from $M$ tangentially bonded
hard-sphere segments of diameter $\sigma_{PS}$. The hard-sphere
monomers forming the chains are freely jointed i.e. they can
adopt any configuration as long as it is free of the intermolecular 
and intramolecular overlap. Within the DFT approach the grand potential
of the system, $\Omega$ is a functional of the local densities of polymers,
$\rho_{P}({\bf R})$ and colloids, $\rho_{C}({\bf r})$
\begin{eqnarray}\label{eq:1}
\lefteqn{\Omega[\rho_P({\bf R}),\rho_{C}({\bf r})]=
F[\rho_P({\bf R}),\rho_{C}({\bf r})]+}\nonumber\\
& &\int\!\!d{\bf R}\rho_{P}({\bf R})(V_{ext}^{(P)}({\bf R})-\mu_{P})\nonumber\\
&+&\int\!\!d{\bf r}\rho_{C}({\bf r})
(V_{ext}^{(C)}({\bf r})-\mu_{C})\;.
\end{eqnarray}
In the above $V_{ext}^{(C)}({\bf r})$, $\mu_C$,  $V_{ext}^{(P)}({\bf R})$
and $\mu_P$ are the external and the chemical potentials for
colloids and polymers, respectively. 
${\bf R}\equiv ({\bf r}_{1}, {\bf r}_{2}, \cdots, {\bf r}_{M})$ denotes
a set of monomer coordinates.
\begin{figure}[t]
\includegraphics[clip,width=8cm]{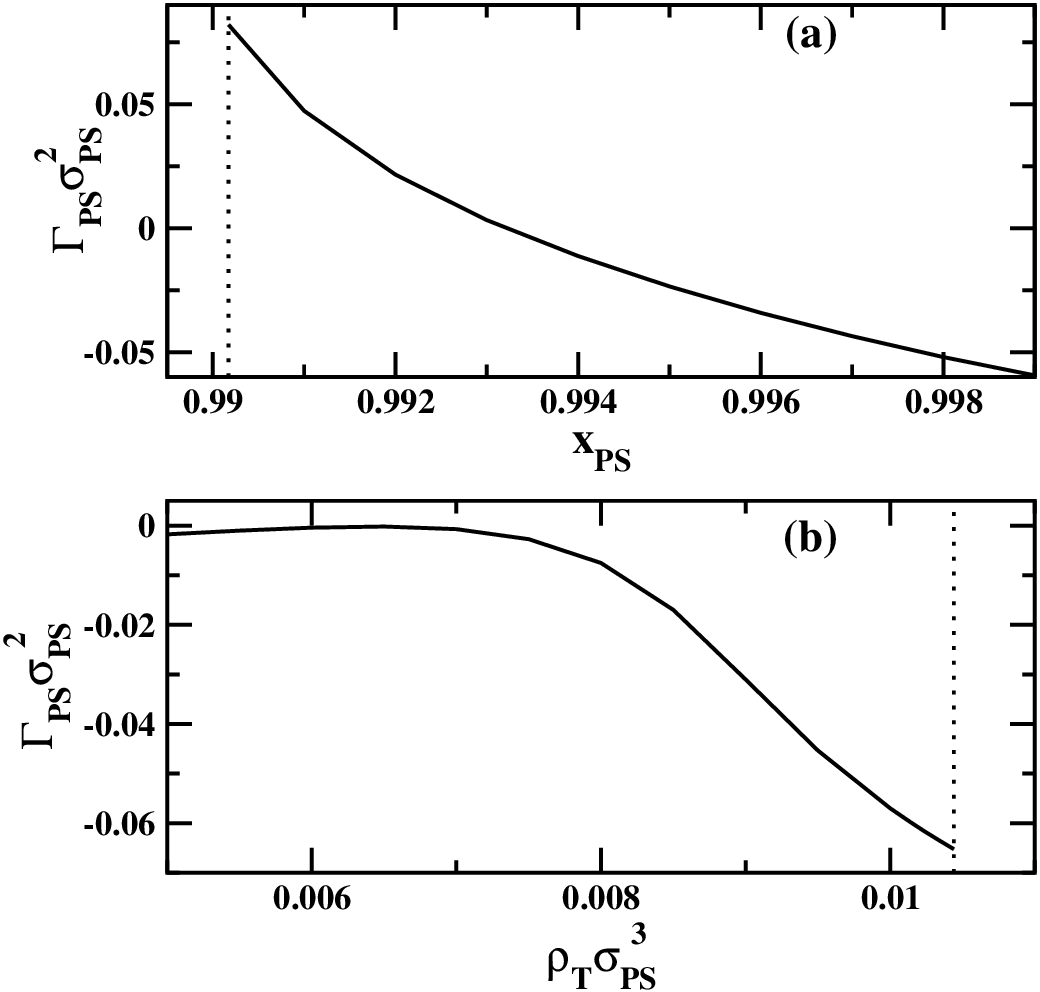}
\caption{\label{fig:2}
The excess polymer segment adsorption isotherm $\Gamma_{PS}$ for
a colloid-polymer mixture with $d=10$ and $M=100$, calculated along path I from Fig.~\ref{fig:1} (upper panel)
and along path II from Fig.~\ref{fig:1} (lower panel).
The dotted lines denote bulk coexistence.
}
\end{figure}
The free energy of the system $F$ is a sum of
the ideal and excess contributions, $F=F_{id}+F_{ex}$.
The ideal part of the free energy is known exactly,
$\beta F_{id}[\rho_P({\bf R}),\rho_{C}({\bf r})]=
\beta\int\!\!d{\bf R}\rho_{P}({\bf R})V_{b}({\bf R})
+\int\!\!d{\bf R}\rho_{P}({\bf R})[\ln(\rho_{P}({\bf R}))-1]
+\int\!\!d{\bf r} \rho_{C}({\bf r})[\ln(\rho_{C}({\bf r}))-1]$.
The total bonding potential, $V_{b}({\bf R})$, is represented as a sum
of the bonding potentials $v_b$ between the monomers,
$V_{b}({\bf R})=\sum_{i=1}^{M-1}v_b(|{\bf r}_{i+1}-{\bf r}_{i}|)$, and satisfies
$\exp [-\beta V_{b}({\bf R})]\propto
\prod_{i=1}^{M-1}\delta(|{\bf r}_{i+1}-{\bf r}_{i}|-\sigma_{PS})$.
Furthermore, it is assumed that the excess free energy is a functional
of the local density of colloids and average segment local density defined as
$\rho_{PS}({\bf r})=\sum_{i=1}^{M}\rho_{PS,i}({\bf r})=\sum_{i=1}^{M}
\int\!\!d{\bf R}\delta({\bf r}-{\bf r}_i)\rho_{P}({\bf R})$.

\begin{figure}[t]
\includegraphics[clip,width=8cm]{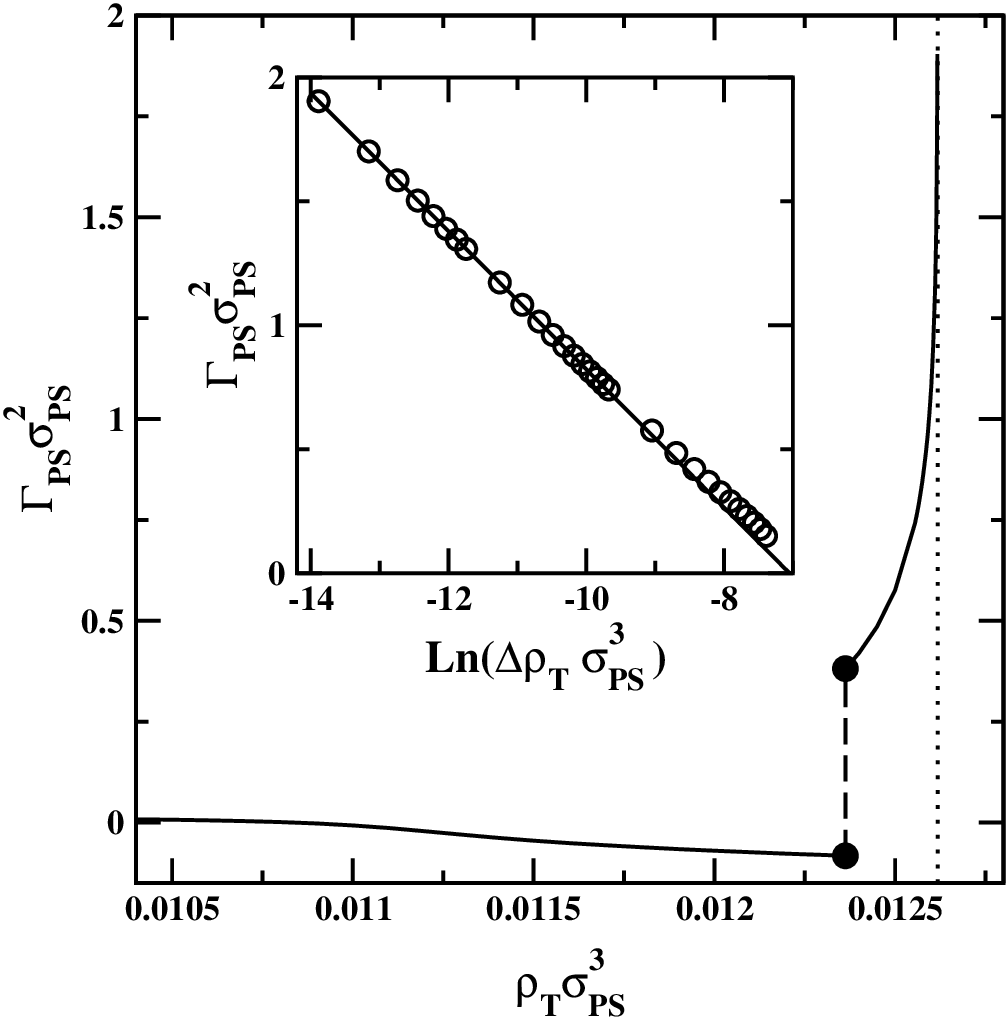}
\caption{\label{fig:3}
The excess polymer segment adsorption isotherm $\Gamma_{PS}$ for
a colloid-polymer mixture with $d=10$ and $M=100$, calculated along a path of constant polymer segment
packing fraction $x_{PS}=$0.944 727 94. The dashed line marks the prewetting transition,
while the dotted line denotes bulk coexistence. The inset illustrates the nature of the
divergence of the excess adsorption (open circles) upon approaching bulk coexistence.
$\Gamma_{PS}$ increases  linearly as a function of the logarithm of the undersaturation $\Delta\rho_T$.
The straight solid line is a guide to the eye.
}
\end{figure}

Within the approach of Yu and Wu \cite{Yu02} the excess free energy is
a volume integral over the excess free energy density, 
$F_{ex}=\int d{\bf r}\{\Phi_{HS}+\Phi_{P}\}$ where $\Phi_{HS}$ is the
excess free energy density of the reference mixture of hard spheres
and $\Phi_{P}$ is a perturbation contribution due to the chain connectivity.
$\Phi_{HS}$ is evaluated using the elegant and inspiring White Bear version \cite{Roth02,Yu02b} of
Rosenfeld's fundamental measure theory (FMT) \cite{Rosenfeld89}. Likewise, FMT-style weighted
densities are used for the polymer contribution, $\Phi_{P}$.
We refer the reader to earlier papers for explicit formulae \cite{Roth02,Yu02,Yu02b}.
The equilibrium density profiles were found from the condition
\begin{equation}\label{eq:2}
\frac{\delta\Omega [\rho_P({\bf R}),\rho_{C}({\bf r})]}
{\delta \rho_P(\mathbf{R})}=
\frac{\delta\Omega [\rho_P({\bf R}),\rho_{C}({\bf r})]}
{\delta \rho_C({\bf r})}=0 \;.
\end{equation}

\section{Results}
\label{sec:results}

Paricaud {\it et al}. \cite{Paricaud03} shown that within the Wertheim approach a mixture of colloids and
excluded volume polymers undergoes a demixing transition into the colloid-rich (polymer-poor)
and colloid-poor (polymer-rich) phases. The properties of the fluid-fluid interface
resulting from this theory were reported in Ref.~\cite{Bryk05}. In the present paper we investigate the wettability
of hard structureless walls by such mixtures. The calculations were carried out
for constant size ratio $d=\sigma_C/\sigma_{PS}=10$ and for chain lengths $M=80$, 100, 120 and 200.
This corresponds to $q=0.894$, 1, 1.095 and 1.414, 
and in the above we assumed that $R_g=a_P M^\nu\sigma_{PS}$ with $a_P\approx0.5$
and $\nu=0.5$\cite{Paricaud03,Dautenhahn94,Lue00,Vega00}.
Thus we consider both regimes $q<1$ and $q>1$.
In Fig.~\ref{fig:1} we recall the bulk phase diagram for the system with $M=100$ plotted in the
polymer segment mole fraction, $x_{PS}=\rho_{PS}^{(b)}/(\rho_{PS}^{(b)}+\rho_C^{(b)})$,
- total density, $\rho_T=\rho_{PS}^{(b)}+\rho_C^{(b)}$, plane.
This representation proved to be useful in carrying out numerical calculations presented below.
If we disregarded the chain connectivity 
$x_{PS}$ would correspond to the mole fraction of the small spheres in a mixture of big and small spheres,
while $\rho_{PS}^{(b)}$ and $\rho_C^{(b)}$ 
would correspond to the bulk number densities of small and big hard spheres, respectively. 
The bulk densities $\rho_{PS}^{(b)}$ and $\rho_C^{(b)}$ serve as an input to the DFT calculations.

We note that the critical polymer segment mole fraction for the system depicted in Fig.~\ref{fig:1}
is 0.976 763 34, thus subcritical state points with lower mole fractions correspond to the polymer-poor (colloid-rich) side
of the phase diagram while state points with $x_{PS}$ higher than this value correspond to the the polymer-rich
(colloid-poor) side. The numerical procedure of the evaluation of the surface phase diagrams relied on
monitoring excess adsorption isotherms, $\Gamma_\alpha=\int d z \rho_\alpha(z)-\rho_\alpha^{(b)}$, $\alpha=$C, PS,
and grand potentials (Eq.~\ref{eq:1}) calculated along paths of constant total density (an example of such
path labelled as ``I'' is shown in Fig.~\ref{fig:1}), or along paths
of constant polymer segment mole fraction (the paths marked as ``II'' and ``III''
in Fig.~\ref{fig:1}).

The inset to Fig.~\ref{fig:1} shows a comparison of the binodal evaluated for $M=100$
resulting from the Werthiem TPT1 description with the binodal resulting from the free-volume theory
for the AOV model \cite{Lekkerkerker92,Brader03} for $q=1$. 
The diagrams are plotted in the colloid-polymer packing fraction representation.
For such comparison to be possible one has to transform the microscopic polymer segment density $\rho_{PS}^{(b)}$
into the polymer packing fraction defined as $\eta_P=4/3\pi R_g^3\rho_{PS}^{(b)}/M$. 
When compared with the AOV model the Wertheim TPT1 approach leads to a shift
of the colloid and polymer critical packing fractions
similar to
the perturbative treatment of Warren {\it et al}. \cite{Warren95} and
the geometrical-based approach \cite{Schmidt03}.
We note that the polymer packing fractions are too low when compared with experimental
values \cite{Aarts02}. This is in part due to the fact that, as mentioned earlier,
the Wertheim TPT1 theory leads to the incorrect Flory exponent
($\nu=0.5$ rather than $0.588$).

We start the presentation of the results of DFT calculations by
analyzing the colloid-poor side of the bulk phase diagram.
The upper panel of Fig.~\ref{fig:2} shows the excess polymer segment adsorption isotherm
evaluated along path I, i.e. along a path of constant total density.
We find the adsorption isotherm to be smooth and small in magnitude. We have carefully inspected
adsorption isotherms calculated along paths similar to the path I for the system with $M=100$ and other
chain lengths and have always found similar behavior.
Therefore we conclude that within the TPT1 formalism mixtures with the state points located on 
the colloid-poor side of the phase diagram always stays in the partial wetting/drying regime. 
This is the opposite of what is found in the wetting studies of the AOV model 
\cite{Brader02,Dijkstra02,Aarts04b}.

\begin{figure}[t]
\includegraphics[clip,width=8cm]{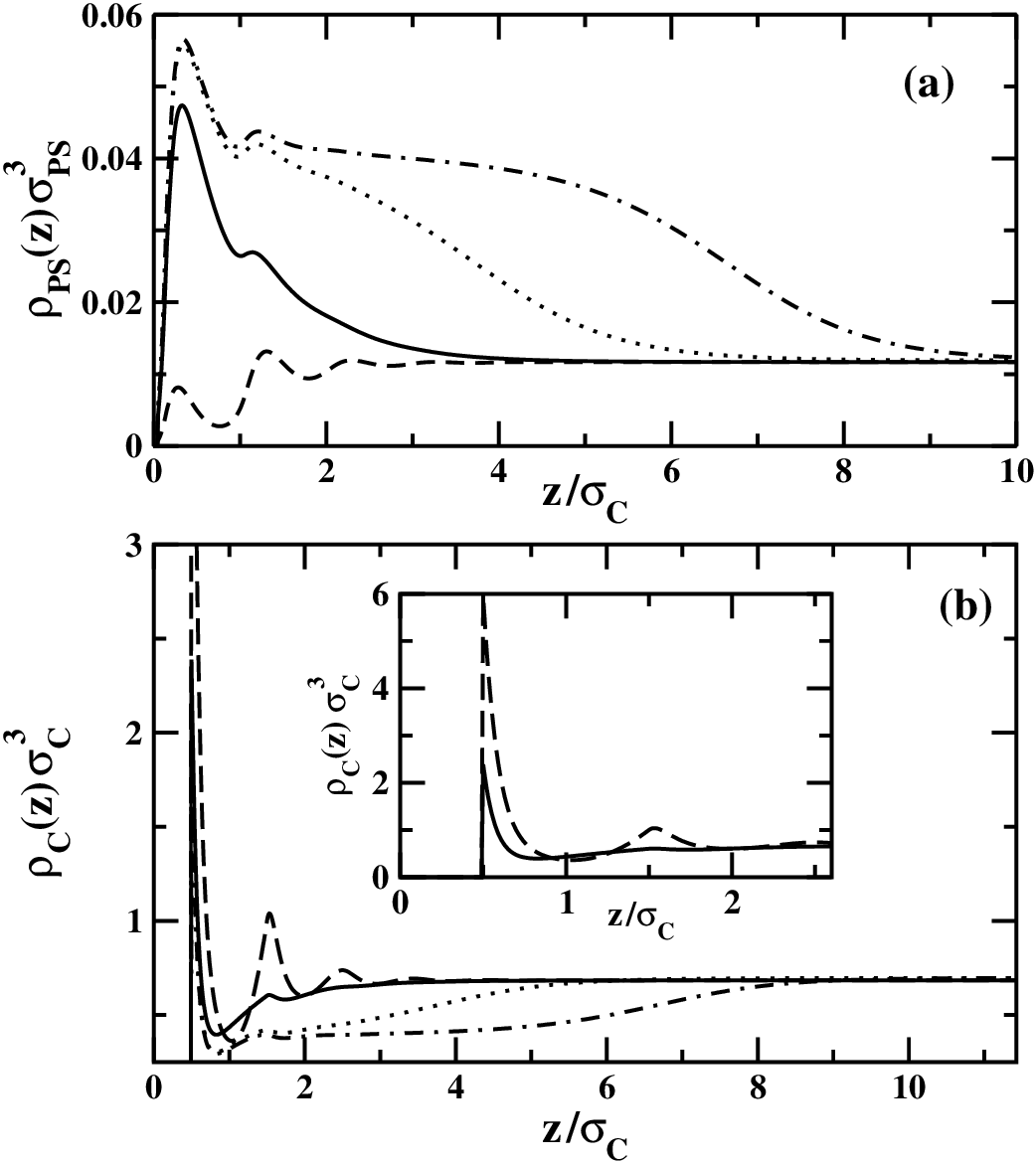}
\caption{\label{fig:4}
Average polymer segment (a) and colloid (b) density profiles for
a colloid-polymer mixture with $d=10$ and $M=100$ near a hard wall.
The profiles are evaluated for constant polymer segment
packing fraction $x_{PS}=$0.944 727 94 and for total bulk densities $\rho_{T}\sigma_{PS}^3=$
0.012 3625  (the dashed and solid lines correspond to the
coexisting thin and thick film phases of the prewetting transition, respectively),
0.0126 (dotted lines), and  0.012 617 (dash-dotted lines).
}
\end{figure}

We turn now to the wetting properties of the state points located on
the polymer-poor side of the phase diagram.
The lower panel of Fig.~\ref{fig:2} shows the excess polymer segment adsorption isotherm
calculated along path II, i.e. along a path of constant polymer segment mole fraction $x_{PS}=$0.928 889 85.
Similar to the path I we find the adsorption isotherm to be smooth, small in magnitude, and
finite at bulk coexistence. However, when paths located closer to the bulk critical point
are considered, a drastic change in the behavior of the inhomogeneous colloid-polymer mixture is observed.
This is illustrated in Fig.~\ref{fig:3}, where we show the adsorption isotherm,
calculated along path III i.e. along a path of constant $x_{PS}=$0.944 727 94.
At lower total bulk densities the excess adsorption is very similar to that evaluated along path II.
However, as $\rho_T$ increases the adsorption isotherm
makes a sudden jump (marked by the dashed line) to a large positive value and diverges upon 
approaching bulk coexistence (denoted by the vertical dotted line). The nature of this divergence
is investigated further in the inset to Fig.~\ref{fig:3}. We find that the excess adsorption
diverges logarithmically and this behavior is characteristic of complete wetting for short range 
forces \cite{Dietrich88}. We therefore conclude that in the TPT1 description the colloid-poor 
demixed phase wets completely the hard wall at coexistence. The wetting transition is of the first
order and is accompanied by the prewetting (or thin-thick film) transition. 

Morphology of the demixed mixture close to the hard wall is investigated further 
in Fig.~\ref{fig:4} where we show examples of the density profiles evaluated at
fixed $x_{PS}=$0.944 727 94 and for several total bulk densities.
The dashed and solid lines correspond to the coexisting thin and thick film profiles 
(these states are marked as black dots in Fig.~\ref{fig:3}). The polymer segment profile
(cf. Fig.~\ref{fig:4}a, dashed line)
corresponding to the thin film phase exhibits oscillations in the direct proximity of the wall, and
the oscillation period approximately equals the colloid diameter. However the excess adsorption is small in
magnitude and negative. The colloid profile
(cf. Fig.~\ref{fig:4}b, dashed line) also exhibits oscillations but
the structure is much more pronounced and the profile exhibits a sharp and high first peak
with a large contact value. This suggests that in the thin film phase the polymers are depleted from the wall,
whereas the colloids are attracted to the wall and form a highly packed first layer.

On the other hand, in the thick film phase of the polymer segment density profile   
(cf. Fig.~\ref{fig:4}a, solid line) one observes  formation of a thick polymer layer next to the wall.
As the total density increases (cf.  Fig.~\ref{fig:4}a, dotted and dash-dotted lines) the polymer-rich
layer grows and extends over the distance of many colloid diameters. The structure of the colloids in the
thick film phase is a reverse of the structure of the polymer. The contact value of the colloid profile
(see the inset to Fig.~\ref{fig:4}b) is reduced by more than a half at the prewetting transition \cite{note1} 
suggesting that the colloids are depleted from the region adjoining the wall. 
The colloid-poor region broadens upon approaching bulk coexistence
(Fig.~\ref{fig:4}b, dotted and dash-dotted lines).

\begin{figure}[t]
\includegraphics[clip,width=8cm]{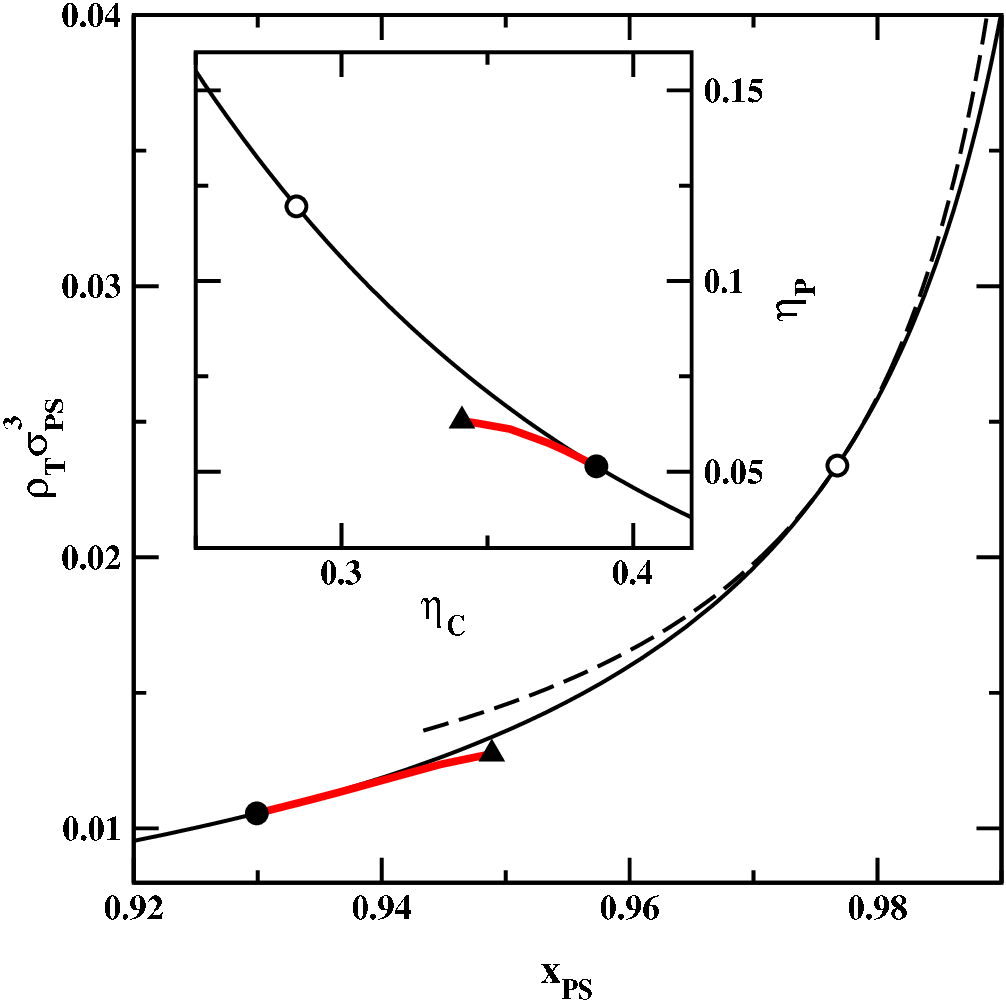}
\caption{\label{fig:5}
Surface phase diagram for a colloid polymer mixture with $d=10$ and $M=100$
in the polymer segment mole fraction-total density plane. The binodal and the spinodal are
denoted by the solid and dashed lines, respectively. The open circle denotes the bulk critical point.
The filled circle denotes the wetting point whereas the triangle denotes the surface critical point.
Thick solid line denotes the prewetting line.
The inset shows the surface phase diagram in the colloid-polymer packing fraction representation.
}
\end{figure}

\begin{figure}[t]
\includegraphics[clip,width=8cm]{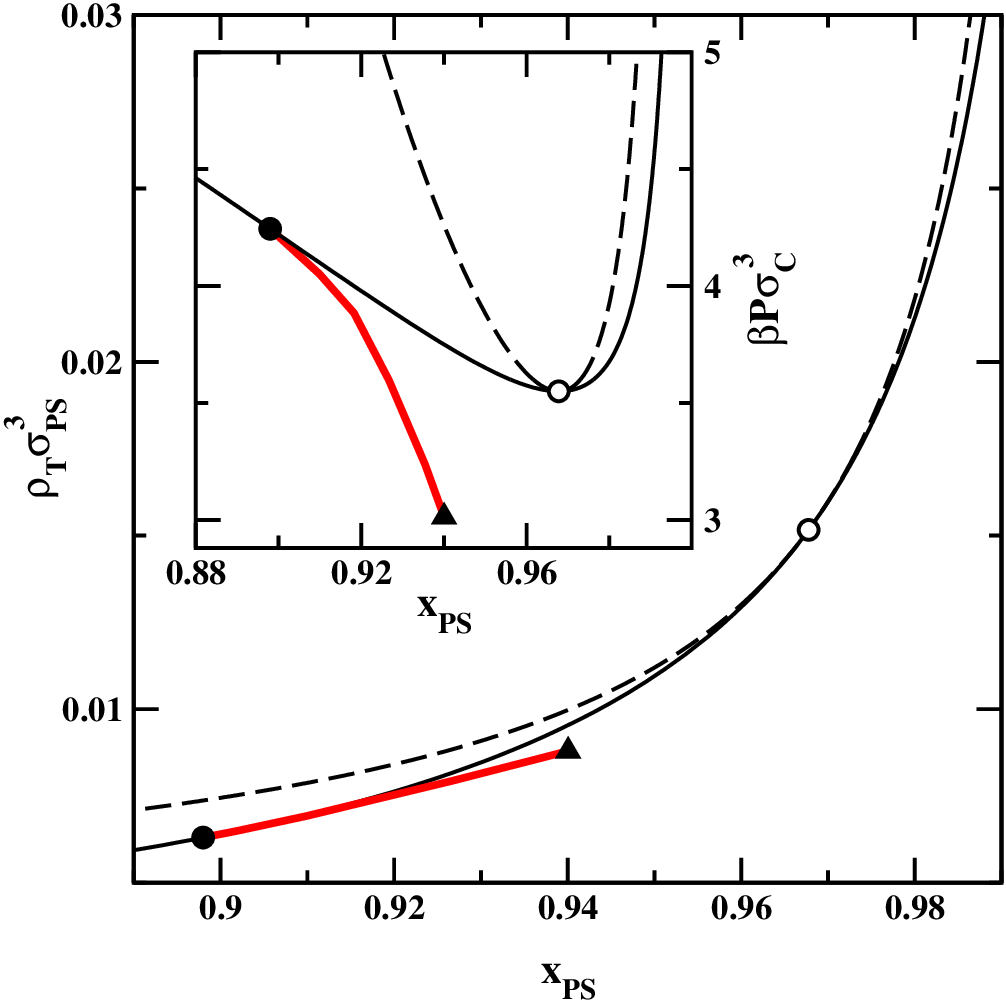}
\caption{\label{fig:6}
Surface phase diagram for a colloid polymer mixture with $d=10$ and $M=200$
in the polymer segment mole fraction-total density plane. The binodal and the spinodal are
denoted by the solid and dashed lines, respectively. The open circle denotes the bulk critical point.
The filled circle denotes the wetting point whereas the triangle denotes the surface critical point.
Thick solid line denotes the prewetting line.
The inset shows the surface phase diagram in the polymer segment mole fraction-pressure representation.
}
\end{figure}

Figures \ref{fig:5} and \ref{fig:6} show the surface phase diagram for colloid-polymer mixtures 
with $M=100$ and $M=200$, respectively.
As in simple fluids the prewetting line tangentially meets the binodal in the wetting transition point
(marked by the filled circle) and ends at the surface critical point (marked by the triangle).
In the polymer segment mole fraction-total density
representation the prewetting line is located very close to the binodal and the surface critical point
is well removed from the bulk critical point (marked by the open circle). However, when 
the colloid-polymer packing fraction (cf. the inset to Fig.~\ref{fig:5})
or the polymer segment mole fraction-pressure representation is chosen (cf. the inset to Fig.~\ref{fig:6}),
the prewetting line is more distant from the binodal. We note that for all chain lengths considered 
the surface critical point pressure is lower
than the pressure of the bulk demixing transition. This curious feature was also found in a recent
study of the wetting behavior of polymer solutions \cite{Forsman05}. However in the present paper we do not find
the lower wetting points and, despite considerable effort, we do not find
any hints for the layering transitions in the partial wetting regime. A likely explanation
to the last observation is that Forsman and Woodward \cite{Forsman05} considered a mixture of polymers and solvent molecules
with diameter ratio $d<1$. Consequently, the individual polymer segments were effectively attracted to the wall due to
the depletion forces induced by the overlap of the excluded volumes of the polymer segments and the wall.
In the present study this effect does not come into play.

In a recent study of wettability of solid surfaces by short-chain fluids \cite{Bryk04}
it was suggested that the wetting
temperature for a given surface, when divided by the critical temperature, is almost independent on $M$ 
for longer chains. Although in the athermal systems considered in the present paper the temperature
does not have any significant role (is a simple scaling parameter), other parameters like the reciprocal pressure 
play the role similar to the temperature in thermal systems. Therefore the wetting point pressure rescaled by the
critical pressure $\tilde{P}_w=P^{(cr)}/P_w$ should barely depend on the chain length.
The estimated wetting point pressure $\beta P_w\sigma_C^3$ for systems $M=80$, 100, 120, and 200 is 
9.75, 7.76, 6.51, and  4.25, respectively.
The corresponding $\tilde{P}_w$ thus are 0.814, 0.818, 0.824, and 0.835. This agrees nicely
with the prediction \cite{Bryk04}.

\section{Discussion}

That the rescaled $\tilde{P}_w$ barely depends on $M$ has a simple explanation.
As demonstrated in Refs.~\cite{Paricaud03,Bryk05}, in the limit of an infinite chain length $M\to\infty$, and for fixed $d$,
the bulk critical pressure (and also the colloid critical packing fraction)
tend to a finite, nonzero value with the leading order correction $\sim {\cal O}(1/\sqrt M)$.
This is the so-called ``protein limit'' of colloid-polymer mixtures \cite{Bolhuis03}.
But then the wetting point pressure also should tend to a constant in that limit.
Therefore the rescaled wetting point pressure should barely depend on M for longer chains.
We also note that the systematic deviations of $\tilde{P}_w$ from a constant suggest that
the leading order chain-length dependent correction to $P_w$ is of different order
than that for the bulk critical point. However the precise form of these corrections
is rather difficult to estimate and we leave this problem for future work.

Wetting in the AOV model for $q<1$ can be readily explained by invoking  
depletion interactions. The overlap of excluded volumes of a colloid and a wall is
larger than the overlap between two colloidal particles and this gives rise
to net colloid-wall attraction. Consequently the colloid-rich phase favors the hard wall.
Our results are in opposition to the AOV model results and
show some features that are characteristic of wetting in athermal polymer solutions.

The suggestion that the polymer-rich phase should wet the hard wall seems
reasonable when $q>1$. However, we find this behavior also for $q<1$ i.e. for $d=10$, $M=80$
(in order to be completely assured that
our calculations are representative of the TPT1 description and are not a result of
e.g. an artifact connected with the incorrect scaling of the polymer gyration radius,
we checked that even for colloid-to-polymer-segment size ratios as large as $d=20$ and for 
chains as short as $M=40$ the polymer-rich phase wets completely the hard wall).
This is at variance with experimental results.
We connect this discrepancy with the inaccurate description of the bulk phase behavior resulting from the TPT1 theory,
and identify two possible sources of errors.
The first issue is connected with the incorrect scaling of the gyration radius 
and of the polymer second virial coefficient with $M$. The consequence is that the dilute and
semidilute regimes of long chains are not described accurately. 
This problem has already been recognized in the literature. Vega {\it et al.} 
proposed a simple correction that significantly improves the accuracy
of the TPT1 description in the low density limit \cite{Vega00}. A similar treatment, which combines
the field-theoretic approach with liquid-state theory, was proposed by  Lue \cite{Lue00}.
However, at present it is not clear how these ideas can be extended 
to mixtures.
To explain the second deficiency in the TPT1 description of colloid-polymer mixtures we recall
that the Wertheim approach, as any other perturbation theory,
relies on the good description of the reference system. In this case it is a mixture
of big and small hard spheres and we used the usual Boublik-Mansoori-Carnahan-Starling-Leland (BMCSL)
equation of state \cite{Boublik70,Mansoori71}. However, it was argued  \cite{Dijkstra99} that
for highly asymmetric hard sphere mixtures
a fluid-fluid phase separation occurs, which is metastable to a broad fluid-solid phase transition.
Consequently the BMCSL equation of state is not very accurate. 
The reference system in the TPT1 approach to colloid-polymer mixtures
consists of a very low packing fraction of small spheres, therefore the potential
impact of this problem seems to be moderate.

\section{Conclusions}
\label{sec:conclusions}
We have used a microscopic density functional theory based on Wertheim's first
order thermodynamic perturbation theory to study the wetting behavior of
athermal mixtures of colloids and excluded-volume polymers.
In opposition to the wetting behavior in the AOV model we have found the polymer-rich phase to wet a hard wall.
The wetting transition is of first order and is accompanied by the prewetting transition. We have not
found any hints for the layering transitions in the partial wetting regime.
The rescaled wetting point pressure is very similar for mixtures with different chain lengths
and the surface critical point pressure is lower than the pressure of the bulk demixing transition.
We have found our results to resemble the wetting behavior in athermal polymer solutions.

Our considerations show that it is certainly a challenge to devise a monomer-resolved
theory for colloid-polymer mixtures that is accurate in both the 'colloid', and 'protein' limits. 
However, we feel that the Wertheim approach could be a good candidate but
this theory should incorporate the correct scaling behavior in the dilute polymer regime
and an accurate description of the reference system.

\begin{acknowledgments}
This work has been supported by KBN of Poland under the Grant
1P03B03326 (years 2004-2006).
\end{acknowledgments}


\begin{thebibliography}{99}
\bibitem{note1} The sum of the contact values of the colloid and the average polymer segment
density profiles is the same for the coexisting thin and thick film phases as dictated by
the appropriate sum rule.
\end{thebibliography}
\end{document}